\documentclass{cernyrep}

\usepackage{graphicx}				
\usepackage{color}					
\usepackage{amssymb}
\newcounter{query}

\renewcommand{\rmd}{{\rm d}}

\title{Introduction to Particle Accelerators  and their Limitations}

\author{B.J.~Holzer}
\institute{CERN, Geneva, Switzerland}

\begin{document}
\maketitle

\begin{abstract}
The paper gives an overview of the principles of particle accelerators and their historical development. After introducing the basic concepts, the main emphasis is on sketching the layout of modern storage rings and discussing their limitations in terms of energy and machine performance. Examples of existing machines, among them the Large Hadron Collider (LHC) at CERN, demonstrate the basic principles of and the technical and physical limits that we face in the design and operation of  particle colliders. The push for ever higher beam energies motivates the design of future colliders as well as the development of more efficient acceleration techniques.
\end{abstract}


\section {Introduction}
The study of matter, from initial theories about the structure of the atom to the discovery of the nucleus and, subsequently, of a variety of particles and their interactions, has been summarized in a scientific picture often called the `standard model'; along its way it has driven the development of powerful tools to create the particle beams that are needed to analyse the detailed structure of matter. The largest accelerator to date, the proton--proton Large Hadron Collider (LHC) at CERN in Geneva, operates with an energy per beam of 7~TeV, which corresponds to an available centre-of-mass energy of $E_{\rm cm}=14$~TeV. The LHC is part of a long tradition of technical and physical progress in creating particle beams, accelerating them, and achieving successful collision with micrometre beam sizes.
This article gives a basic introduction to the physics of particle accelerators and discusses some of their limitations. The author has arbitrarily selected ten limitations to focus on, although in fact there are many more that the reader may find in other publications and which could be studied further (and some of them overcome, hopefully).

\subsection {Limit I: The geography}
\begin{figure}[ht]
\begin{center}
\includegraphics[width=0.40\columnwidth]{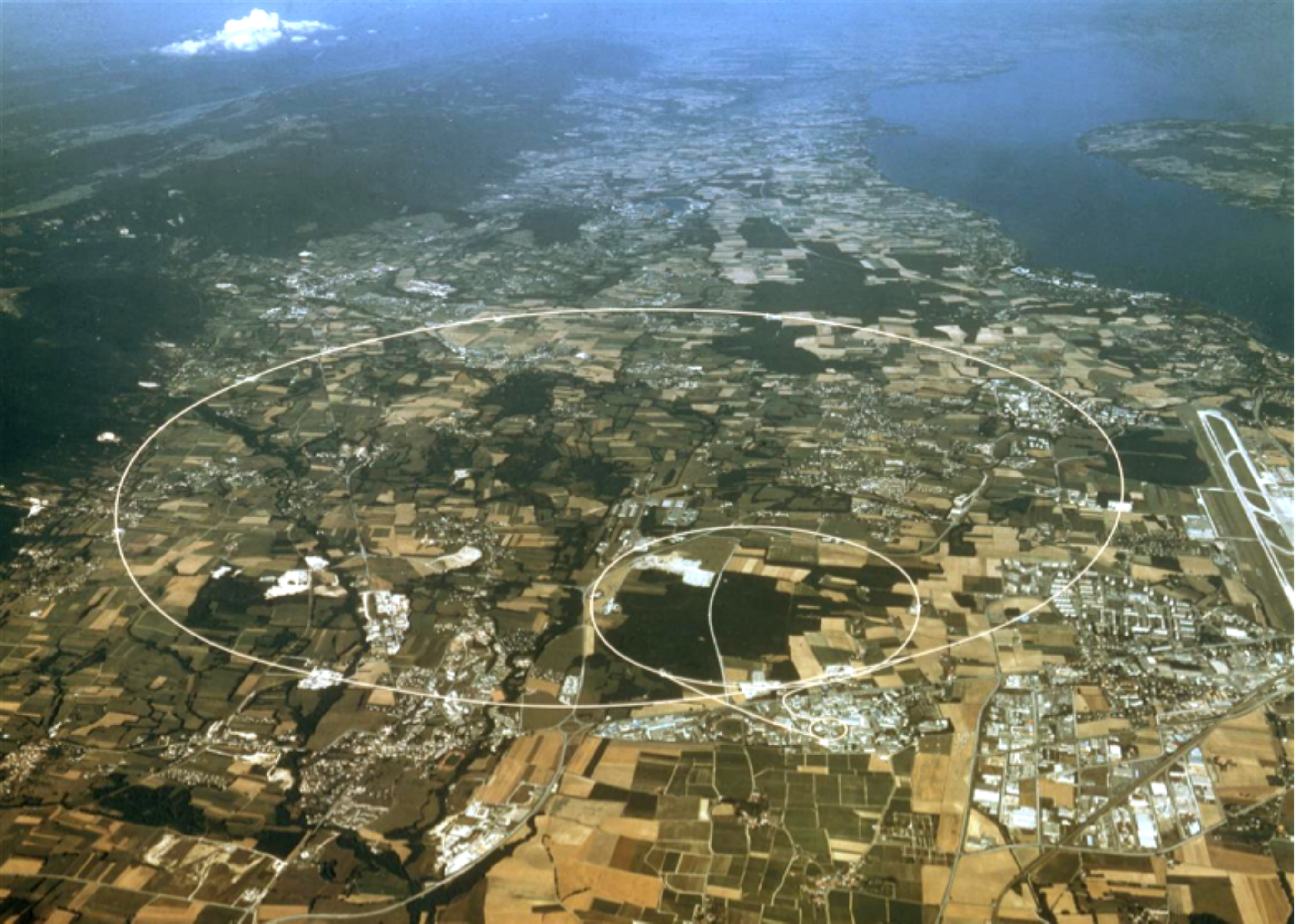}
\caption{The proton--proton collider LHC at CERN, Geneva}
\label{lhc}
\end{center}
\end{figure}

For the fun of it, let us start at the end of the line: the largest accelerators and the fact that beyond physical and technical limits there is a serious boundary condition---the landscape.  For a given technology, pushing the particle energy of a storage ring to higher and higher values will necessitate  larger and larger machines, and we may suddenly encounter the problem that our device no longer fits in the garage at our institute or, as shown in Fig.~\ref{lhc}, not even in the entire region surrounding our facility. As regards LHC \cite{LHC}, the largest storage ring at present, and the Geneva region, the space between Lake Geneva and the Jura mountains defined the size of the tunnel for the present LHC. As a consequence, the maximum feasible beam energy available for high-energy physics experiments is determined by the geographical boundary conditions of the Geneva countryside. Certainly, there are extremely high particle energies in cosmic rays, but you will agree that the accelerators driving these are also much much larger!

\begin{figure}[h!]
\begin{center}
\includegraphics[width=0.5\columnwidth]{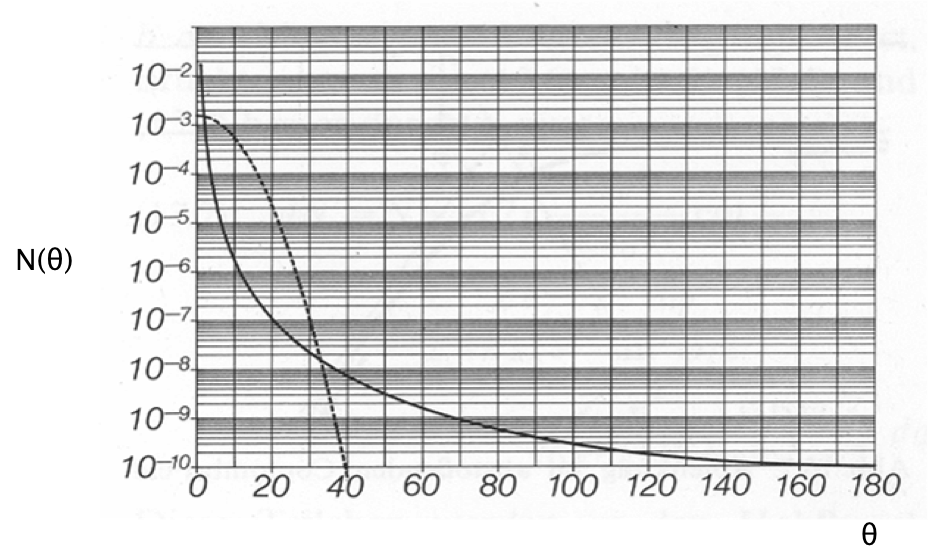}
\caption{Comparison of the Thomson model of the atom and the results of Rutherford's experiment;
the number of scattering events of alpha particles scattered at the gold foil is plotted as a function of the measured (or predicted) scattering angle.}
\label{Rutherford}
\end{center}
\end{figure}

Before we discuss the high-energy frontier machines, let us take a brief look at the path paved by the ingenious scientific developments dating back to the discovery of the nucleus by Ernest Rutherford.  Figure~\ref{Rutherford} (taken from \cite{Gerthsen})
shows a comparison of the scattering events, plotted as a function of the scattering angle, predicted by the Thomson model of the atom and the experimental results obtained by Rutherford: the discrepancy between the model, which assumes equally distributed charges in matter, and the observed data is evident. As a consequence, the concept of the nucleus was born.

Using alpha particles on the level of MeV is not ideal for precise, triggerable and healthy experiments. So Rutherford discussed with two colleagues, Cockcroft and Walton, the possibility of using artificially accelerated particles. Based on this idea, within only four years Cockcroft and Walton invented the first particle accelerator ever built, and in 1932 they gave the first demonstration of the splitting of a nucleus (lithium) by using a 400~keV proton beam.

\begin{figure}[ht]
\begin{center}
\includegraphics[width=0.3\columnwidth]{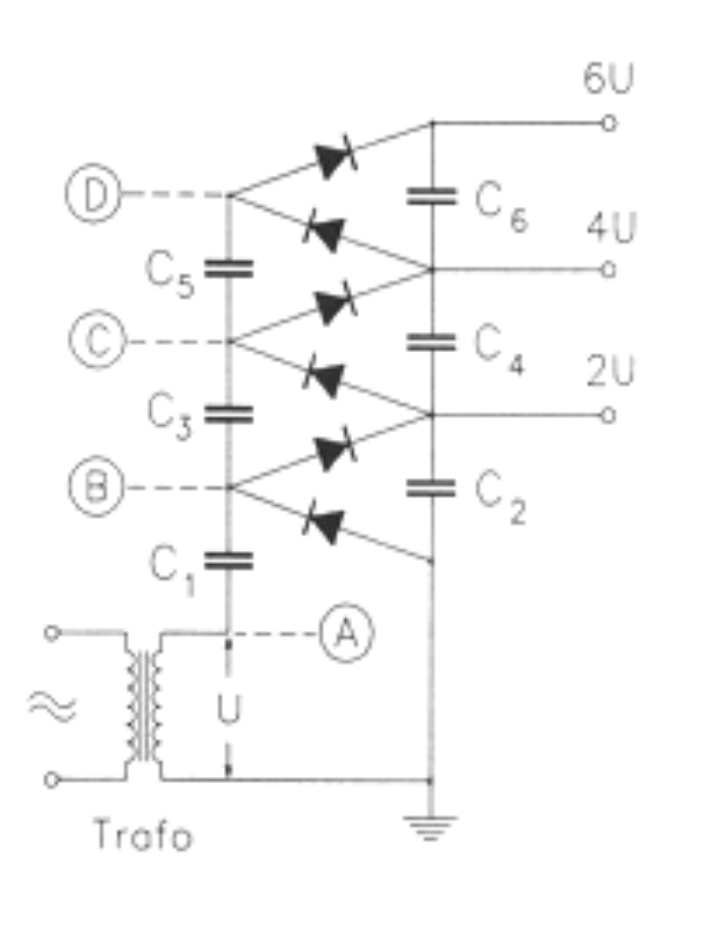}
\caption{Schematic layout of the rectifier circuit used by Cockcroft and Walton to generate the high DC voltage needed for their accelerator. %
}
 \label{Cockcroft_schema}
\end{center}
\end{figure}

\begin{figure}[hb]
\begin{center}
\includegraphics[width=0.4\columnwidth]{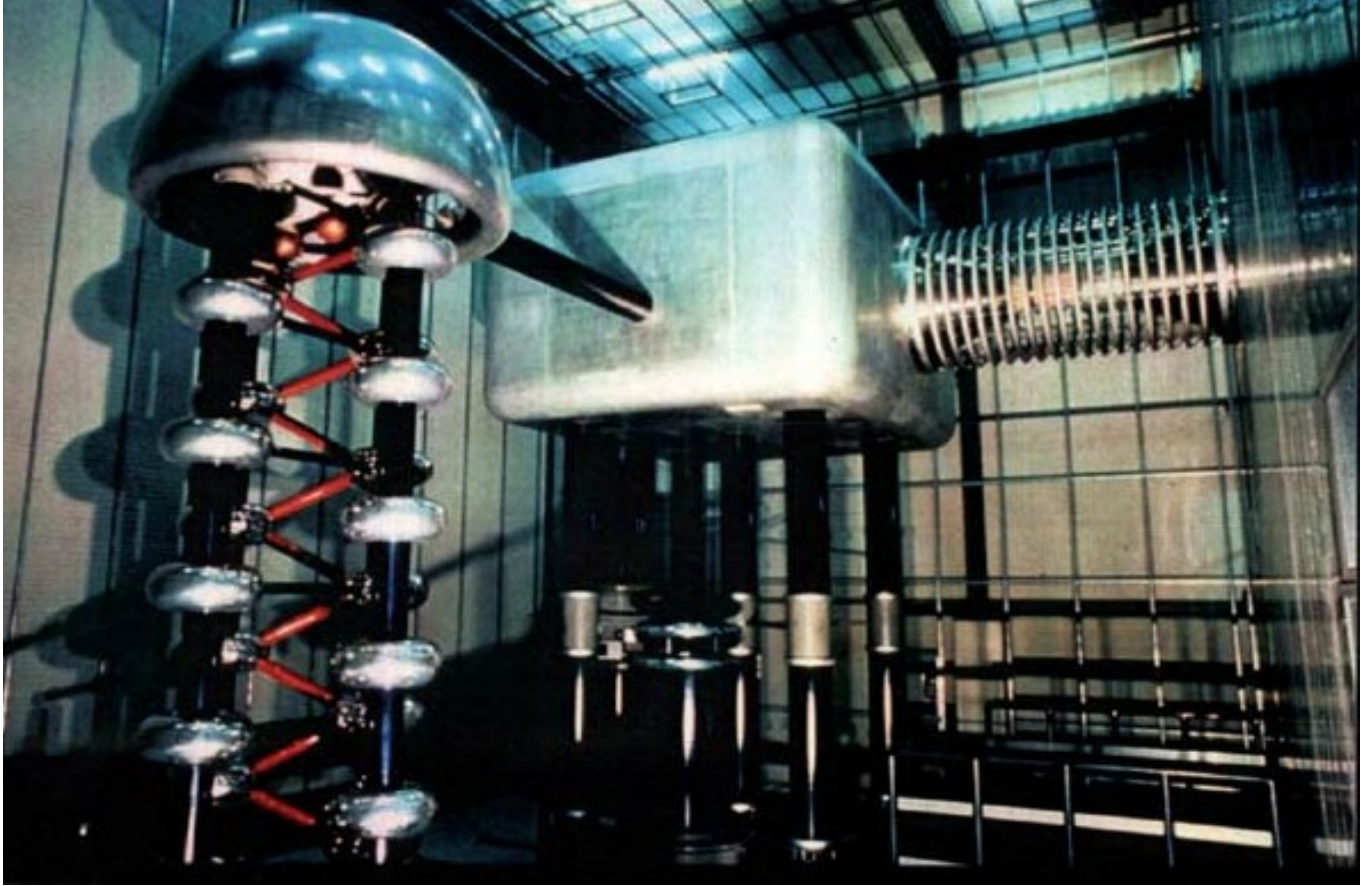}
\caption{A Cockcroft--Walton generator that was used at CERN as a pre-accelerator for the proton beams; the device has since been replaced by the more compact and efficient RFQ technique.
}
 \label{Cockcroft_photo}
\end{center}
\end{figure}

Their acceleration mechanism was based on a rectifier or Greinacher circuit, consisting of a number of diodes and capacitors that transformed a relatively small AC voltage to a DC potential which corresponds, depending on the number of diode/capacitor units used, to a multiple of the applied basic potential.
The particle source was a standard hydrogen discharge source connected to the high-voltage part of the system, and the particle beam was accelerated to ground potential, hitting the lithium target \cite{Cockcroft}.
A schematic view of the mechanism is shown in Fig.~\ref{Cockcroft_schema} (for details see, for instance, Ref.~\cite{WiIle}), and a photograph of such a device which has been used at CERN for many years is presented in Fig.~\ref{Cockcroft_photo}.
Cockcroft and Walton were awarded the Nobel prize for their invention.

\subsection {Limit II: Voltage breakdown in DC accelerators}
\begin{figure}[h!]
\begin{center}
\includegraphics[width=0.35\columnwidth]{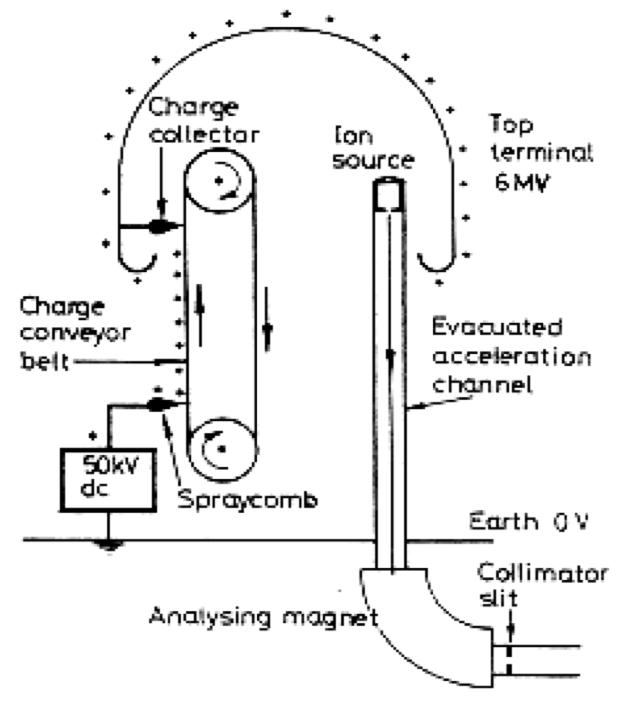}
\caption{Schematic design of a Van de Graaff accelerator%
}
 \label{vdg_schema}
\end{center}
\end{figure}

In parallel to Cockcroft and Walton, but based on a completely different technique, another type of DC accelerator had been invented: Van de Graaff designed a DC accelerator \cite{vdGraaf} that used a mechanical transport system to carry charges, sprayed on a belt or chain, to a high-voltage terminal; see Fig.~\ref{vdg_schema}, taken from Ref.~\cite{Bryant}.

\begin{figure}[h]
\begin{center}
\includegraphics[width=0.55\columnwidth]{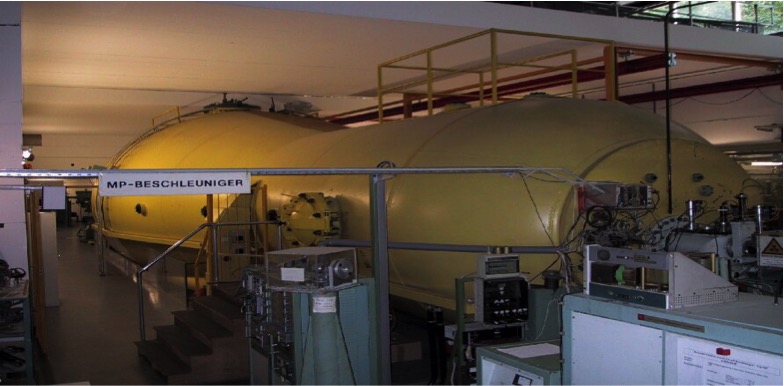}
\caption{A typical example of a tandem Van de Graaff accelerator; these are very reliable machines for precise measurements in atomic and nuclear physics. (Photo courtesy of the Max Planck Institute for Nuclear Physics, Heidelberg.)%
}
 \label{vdg_photo}
\end{center}
\end{figure}

In general these machines can reach higher voltages than the Cockcroft--Walton devices, but they are more limited in terms of particle intensity.
Common to all DC accelerators is the limitation on the achievable beam energy due to high-voltage breakdown effects (discharges). Without using an insulating gas (SF$_6$ in most cases), electric fields will be limited to about 1~MV\,m$^{-1}$, and even with the most sophisticated devices, like the one in Fig.~\ref{vdg_photo}, acceleration voltages on the order of MV cannot be overcome.
In fact, the example in Fig.~\ref{vdg_photo} shows an approach that has been applied in a number of situations: injecting a negative ion beam (even H$^{-}$ is used) and stripping the ions in the middle of the high-voltage terminal allows one to profit from the potential difference twice and thus to make another step of gain in beam energy.

\begin{figure}[h]
\begin{center}
\includegraphics[width=0.7\columnwidth]{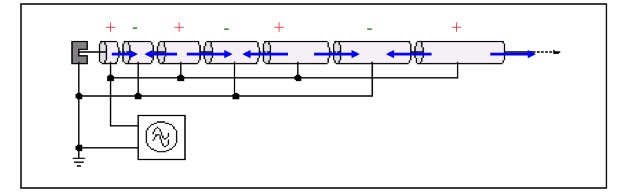}
\caption{Schematic view of the Wider{\o}e principle as a fundamental concept for AC (or RF) acceleration%
}
\label{Wideroe_schema}
\end{center}
\end{figure}

Given the obvious limitations of the  DC machines described above, the next step forward is natural. In 1928, Wider{\o}e developed the concept of a AC accelerator. Instead of rectifying the AC voltage, he connected a series of acceleration electrodes in an alternating manner to the output of an AC supply. The schematic layout is shown in Fig.~\ref{Wideroe_schema} where, for a instant in time, the direction of the electric field is indicated.
In principle, this device can produce step by step a multiple of the acceleration voltage, as long as for the negative half-wave of the AC voltage the particles are shielded from the decelerating field. The energy gain after the $n$th step is therefore
\begin{equation}
E_n=n\cdot q \cdot U_{0} \cdot \sin(\psi_s),
\end{equation}
where $n$ denotes the acceleration step, $q$ the charge of the particle, $U_{0}$ the applied voltage per gap, and $\psi_s$ the phase between the particle and the changing AC voltage.

\subsection{Limit III: The size of the accelerating structure}
\begin{figure}[h!]
\begin{center}
\includegraphics[width=0.45\columnwidth]{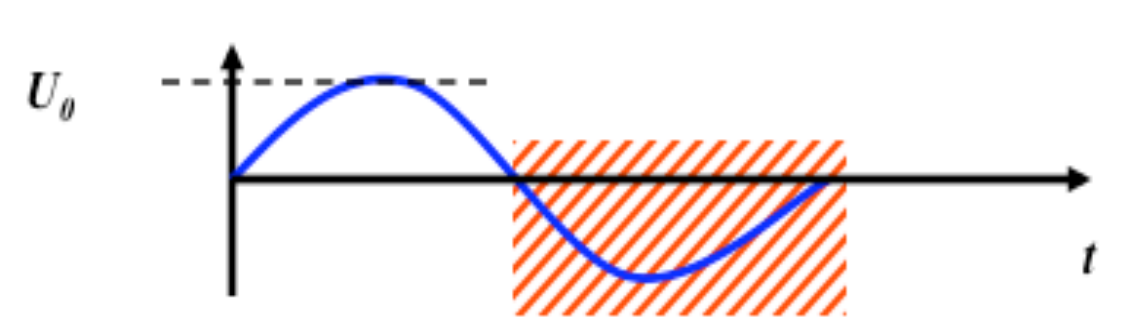}
\caption{The frequency (and hence the period) of the RF system and the particle speed determine the length of the drift tubes in the Wider{\o}e structure.
}
\label{Wiederoe_half_wave}
\end{center}
\end{figure}

A key quantity in such a Wider{\o}e structure is the length of the drift tubes that will protect the particles from the negative half-wave of the sinusoidal AC voltage. For a given frequency of the applied radio-frequency (RF) voltage, the length of the drift tube is defined by the speed of the particle and the duration of the negative half-wave of the sinusoidal voltage, as shown in Fig.~\ref{Wiederoe_half_wave}.

\begin{figure}[h]
\begin{center}
\includegraphics[width=0.5\columnwidth]{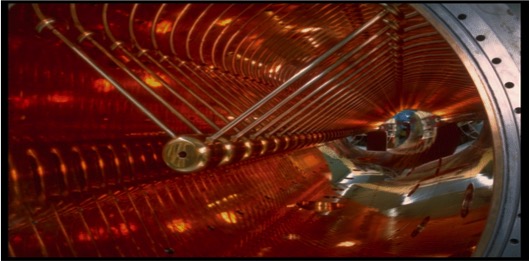}
\caption{Unilac at GSI, Darmstadt; clearly visible are the structure of the drift tubes and their increasing length as a function of the particle energy. %
}
\label{Unilac}
\end{center}
\end{figure}

The time-span of the negative half-wave is defined by the applied frequency, $ \Delta t=\tau_{\rm rf}/2$, so  for the length of the $n$th drift tube we get
\begin{equation}
l_{n}=v_{n} \cdot \frac{\tau_{\rm rf}}{2}.
\end{equation}
Given the kinetic energy of the particle,
\begin{equation}
E_{\rm kin}=\frac {1}{2}mv^{2},
\end{equation}
we obtain directly that
\begin{equation}
l_{n}=\frac {1} {\nu_{\rm rf}} \cdot \sqrt{\frac {nqU_{0}\sin{\psi_{s}}}{2m}},
\end{equation}
which defines the design concept of the machine.
Figure \ref{Unilac} shows a photograph of such a device, the Unilac at the Institute for Heavy Ion Research (GSI) in Darmstadt, Germany.

\noindent Two remarks should be made in this context.
\begin{itemize}
\item The short derivation here is based on the classical approach, and in fact these accelerators are usually optimum for `low-energetic' proton or heavy-ion beams. Typical beam energies (referring to protons) are on the order of  10~MeV; for example, the present Linac~2 at CERN delivers the protons for LHC operation with an energy of 50~MeV, corresponding to a relativistic $\beta$ of 0.31.
\item For higher energies, even in the case of protons or ions, the speed will at some point approach the speed of light, and the length of the drift tubes and hence the dimension of the whole accelerator will reach a size that may no longer be feasible. More advanced ideas are needed in order to keep the machine within reasonable dimensions, and the next natural step in the historical development was to introduce magnetic fields and bend the particle beam into a circle.
\end{itemize}

\section{Pushing for the highest energies: synchrotrons and storage rings}
A significant step forward in achieving high beam energies involves the use of circular structures. In order to apply over and over again the accelerating fields, we try to bend the particles onto a circular path and so bring them back to the RF structure where they will receive the next step-up in energy.
To do this,  we introduce magnetic (or electric) fields that will deflect the particles and keep them on a well-defined orbit during the complete acceleration process. The Lorentz force that acts on a particle will therefore have to compensate exactly the centrifugal force due to the bent orbit.
In general, we can write
\begin{equation}
\mathbf{F} = q \cdot (\mathbf{E}+ \mathbf{v} \times \mathbf{B}).
\end{equation}
For high-energy particle beams, the velocity $v$ is close to the speed of light and so represents a nice amplification factor whenever we apply a magnetic field. As a consequence, it is much more convenient to use magnetic fields for bending and focusing the particles.

Therefore, neglecting electric fields for the moment, we write the Lorentz force and the centrifugal force of the particle on its circular path as
\begin{align}
F_{\rm Lorentz}&=e \cdot v \cdot B,\\
F_{\rm centrifugal}&=\frac{\gamma m_{0} v^{2}}{\rho}.
\end{align}
Assuming an idealized homogeneous dipole oriented along the particle orbit, we define the condition for a perfect circular orbit as equality between these two forces; this yields the following condition for the idealized ring:
\begin{equation}
\frac{p}{e}={B \cdot \rho},
\end{equation}
where we refer to protons and have accordingly set $q = e$.
This condition relates the so-called beam rigidity $B\rho$ to the particle momentum that can be carried in the storage ring, and it will ultimately define, for a given magnetic field of the dipoles, the size of the storage ring.

\begin{figure}[h]
\begin{center}
\includegraphics[width=0.5\columnwidth]{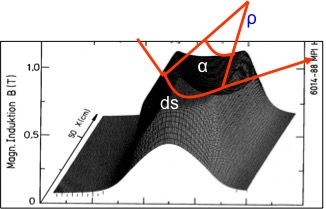}
\caption{Dipole field of a storage ring and schematic path of the particles%
}
\label{TSR_dipole_field}
\end{center}
\end{figure}

In reality, instead of a continuous dipole field the storage ring will be built out of several dipoles, powered in series to define the geometry of the ring.
For a single magnet, the particle trajectory is shown schematically  in Fig.~\ref{TSR_dipole_field}.
In the free space outside the dipole magnet, the particle trajectory follows a straight line. As soon as the particle enters the magnet, it is bent onto a circular path until it leaves the magnet at the other side.

\subsection{Limit IV: The magnetic guide field}
The overall effect of the main bending (or `dipole') magnets in the ring is to define a more or less circular path, which we will call the `design orbit'. By definition, this design orbit has to be a closed loop,  and so  the main dipole magnets in the ring have to define a bending angle of exactly $2 \pi$ overall. If $\alpha$ denotes the bending angle of a  single magnet, then
\begin{equation}
\alpha=\frac{\rmd s}{\rho}=\frac{B\,\rmd s}{B \cdot \rho}.
\end{equation}
We  therefore require that
\begin{equation}
\frac{\int B \,\rmd l}{B \cdot \rho}=2 \pi.
\label{eq_beam_rigidity}
\end{equation}
Thus, a storage ring  is not a `ring' in the true sense of the word but more a polygon, where `poly' means the discrete number of dipole magnets installed in the `ring'.

\begin{figure}[h]
\begin{center}
\includegraphics[width=0.35\columnwidth]{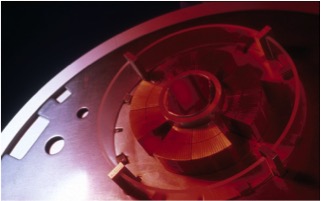}
\caption{Superconducting quadrupole of the LHC storage ring%
}
\label{LHC_quad}
\end{center}
\end{figure}

In the case of the LHC, the dipole field has been pushed to the highest achievable values;
1232 superconducting dipole magnets, each of length 15~m, define the geometry of the ring and, via  Eq.~(\ref{eq_beam_rigidity}),  the maximum momentum for the stored proton beam.
Using the equation given above, for a maximum momentum of $p=7$~TeV/c we obtain a required magnetic field of
\begin{equation}
B=\frac {2 \pi \cdot 7000 \cdot 10^{9}~{\rm eV}}{1232 \cdot 15~{\rm m} \cdot 2.99792 \cdot 10^{8}~{\rm m}\,{\rm s}^{-1}},
\end{equation}
or
\begin{equation}
B=8.33 T,
\end{equation}
 to bend the beams.
For convenience we have expressed the particle momentum in units of GeV/$c$.
Figure~\ref{LHC_quad} shows a photograph of the LHC dipole magnets, built out of superconducting NbTi filaments, which are operated at a temperature of $T=1.9$~K.

\subsection{Focusing properties}
In addition to the main bending magnets that guide the beam onto a closed orbit, focusing fields are needed to keep the particles close together. In modern storage rings and light sources, the particles are kept in the machine for many hours, and a carefully designed focusing structure is needed to maintain the necessary beam size at different locations in the ring.

Following classical mechanics, linear restoring forces are needed, just as in the case of a harmonic pendulum. Quadrupole magnets provide the corresponding property:
they create a magnetic field that depends linearly on the particle amplitude, i.e.\ the distance of the particle from the design orbit:
\begin{equation}
B_{x}=g\cdot y , \qquad
B_{y}=g \cdot x .
\end{equation}
The constant $g$ is called the gradient of the magnetic field and characterizes the focusing strength of the magnetic lens in both transverse planes. For convenience it is (like the dipole field) normalized to the particle momentum. The normalized gradient is denoted by $k$ and defined as
\begin{equation}
k=\frac{g}{p/e}=\frac{g}{B \rho}.
\end{equation}
The technical layout of such a quadrupole is depicted  in Fig.~\ref{LHC_quad}. As in the case of the LHC dipoles, the quadrupole magnet is built in superconducting technology.

\begin{figure}[h]
\begin{center}
\includegraphics[width=0.5\columnwidth]{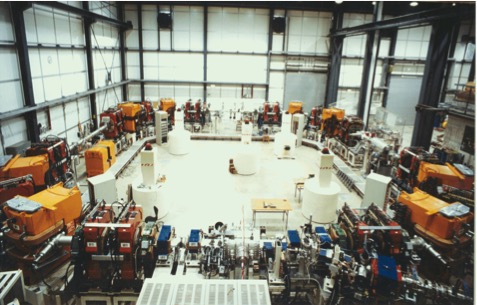}
\caption{TSR storage ring, Heidelberg, as a typical example of a separate-function strong focusing storage ring%
}
\label{TSR_photo}
\end{center}
\end{figure}

Now that we have defined the basic building blocks of a storage ring, we need to arrange them in a so-called magnet lattice and optimize the field strengths in such a way as to obtain the required beam parameters. An example of what such a magnet lattice looks like is given in Fig.~\ref{TSR_photo}. This photograph shows the dipole (orange) and quadrupole (red) magnets in the TSR storage ring in Heidelberg.  Eight dipoles are used to bend the beam in a `circle', and the quadrupole lenses between them provide the focusing to keep the particles within the aperture limits of the vacuum chamber.

A general design principle of modern synchrotrons or storage rings should be pointed out here.
In general, these machines are built following a so-called separate-function scheme: every magnet is designed and optimized for a certain task, such as bending, focusing, chromatic correction, and so on. We separate the magnets in the design according to the job they are supposed to do; only in rare cases a combined-function scheme is chosen, where different magnet properties are combined in one piece of hardware. To express this mathematically, we use the general Taylor expansion of the magnetic field,
\begin{equation}
\frac{B(x)}{p/e}=\frac{1}{\rho}+ k \cdot x + \frac{1}{2 !}mx^{2} + \frac{1}{3 !}nx^{3}+ \cdots.
\end{equation}
Following the arguments above, for the moment we take into account only constant (dipole) or linear  (quadrupole) terms.
The higher-order field contributions will be treated later as (hopefully) small perturbations.

\begin{figure}[h]
\begin{center}
\includegraphics[width=0.4\columnwidth]{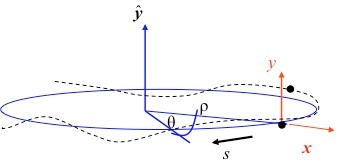}
\caption{Coordinate system used in particle beam dynamics; the longitudinal coordinate $s$ moves  around the ring with the particle considered.%
}
\label{Coordsystem}
\end{center}
\end{figure}

The particles will now follow the `circular' path defined by the dipole fields, and in addition will undergo harmonic oscillations in both transverse planes. The situation is shown schematically in Fig.~\ref{Coordsystem}. An ideal particle will follow the design orbit that is represented  by the circle in the diagram. Any other particle will perform transverse oscillations under the influence of the external focusing fields,  and the amplitude of these oscillations will ultimately define the beam size.

\begin{figure}[h]
\begin{center}
\includegraphics[width=0.27999999999999997\columnwidth]{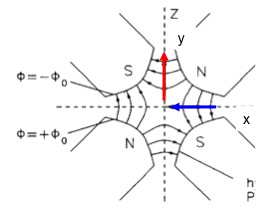}
\caption{Field configuration in a quadrupole magnet and the direction of the focusing and defocusing forces in the horizontal and vertical planes.%
}
\label{Quad_field}
\end{center}
\end{figure}

Unlike a classical harmonic oscillator, however, the equations of motion in the horizontal and vertical planes differ somewhat.
Assuming a horizontal focusing magnet, the equation of motion is
\begin{equation}
x''+x \cdot \left(\frac {1}{\rho^{2}}+k\right)=0,
\end{equation}
where $k$ is the normalized gradient introduced above and the $1/ \rho^{2}$ term represents the so-called weak focusing, which is a property of the bending magnets.
In the vertical plane, on the other hand, due to the orientation of the field lines and by Maxwell's equations, the forces instead have a defocusing effect; also, the weak focusing term disappears:
\begin{equation}
y''- y \cdot k=0.
\end{equation}
The principal problem arising from the different directions of the Lorentz force in the two transverse planes of a quadrupole field  is sketched  in Fig.~\ref{Quad_field}. It is the task of the machine designer to find an adequate solution to this problem and to define a magnet pattern that will provide an overall focusing effect in both transverse planes.

\begin{figure}[h]
\begin{center}
\includegraphics[width=0.5\columnwidth]{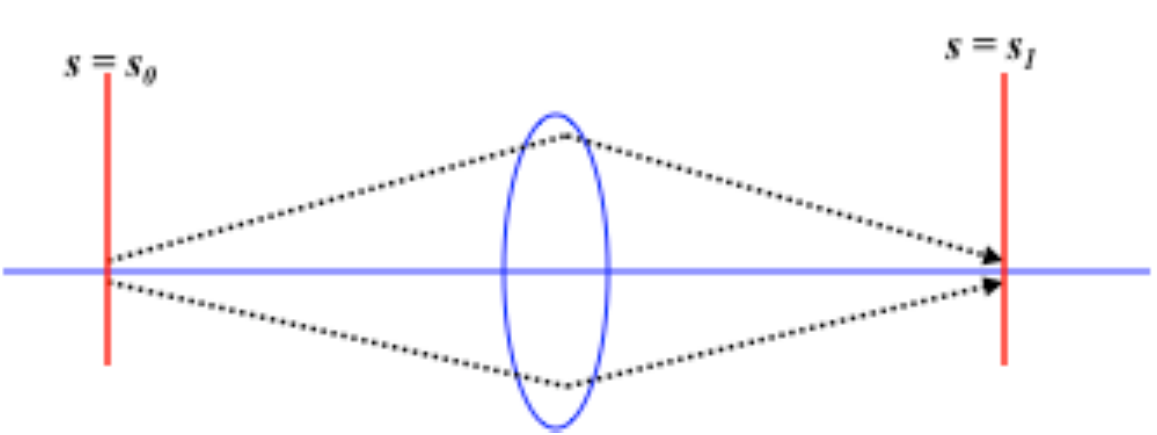}
\caption{Schematic principle of the effect of a focusing quadrupole magnet%
}
\label{Matrix_foc}
\end{center}
\end{figure}

Following closely the example of the classical harmonic oscillator,
we can write down the solutions of the above  equations of motion. For simplicity, we focus on the  horizontal plane; a `focusing' magnet is therefore focusing in this horizontal plane and at the same time defocusing in the vertical plane.
Starting with initial conditions for the particle amplitude $x_{0}$ and angle $x'_{0}$ in front of the magnet element, we obtain the following relations for the trajectory inside the magnet:
\begin{align}
x(s)&=x_{0}\cdot \cos\bigl(\sqrt{{|K|}}\,s\bigr)+x'_{0}\cdot \frac{1}{\sqrt{{|K|}}}\sin\bigl(\sqrt{{|K|}}\,s\bigr), \\
x'(s)&=-x_{0}\cdot \sqrt{{|K|}} \sin\bigl(\sqrt{{|K|}}\,s\bigr)+x'_{0}\cdot \cos\bigl(\sqrt{{|K|}}\,s\bigr).
\end{align}
Here the parameter $K$ combines the quadrupole gradient and the weak focusing effect,
$K=k-\frac{1}{\rho^{2}}$.
Usually these two equations are combined into a more elegant and convenient matrix form:
\begin{equation}
\begin{pmatrix}
x \\
x'
\end{pmatrix}_{s}
=\mathbf{M}_{\rm foc}
\begin{pmatrix}
x \\
x'
\end{pmatrix}_{0},
\end{equation}
where the matrix $\mathbf{M}_{\rm foc}$ contains all the relevant information about the magnet element,
\[
\mathbf{M}_{\rm foc} =
\begin{pmatrix}
\cos(\sqrt{{|K|}}\,s)              &     \frac{1}{\sqrt{{|K|}}} \sin(\sqrt{{|K|}}\,s)  \\
- \sqrt{{|K|}}\sin(\sqrt{{|K|}}\,s)   &   \cos(\sqrt{{|K|}}\,s)
\end{pmatrix}.
\]
Schematically, the situation is visualized in Fig.~\ref{Matrix_foc}.

\begin{figure}[h]
\begin{center}
\includegraphics[width=0.5\columnwidth]{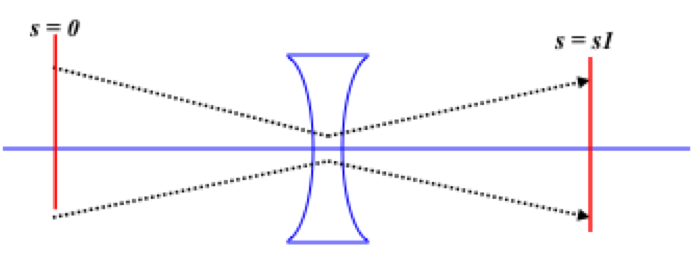}
\caption{Schematic principle of the effect of a defocusing quadrupole magnet%
}
\label{Matrix_defoc}
\end{center}
\end{figure}

In the case of a defocusing magnet, we obtain  analogously that
\begin{equation}
\begin{pmatrix}
x \\
x'
\end{pmatrix}_{s}
=\mathbf{M}_{\rm defoc}
\begin{pmatrix}
x \\
x'
\end{pmatrix}_{0},
\end{equation}
with
\[
\mathbf{M}_{\rm defoc}=
\begin{pmatrix}
\cosh(\sqrt{{|K|}}\,s)              &     \frac{1}{\sqrt{{|K|}}} \sinh(\sqrt{{|K|}}\,s)  \\
\sqrt{{|K|}}\sinh(\sqrt{{|K|}}\,s)   &   \cosh(\sqrt{{|K|}}\,s)
\end{pmatrix};
\]
see Fig.~\ref{Matrix_defoc}.

\begin{figure}[h]
\begin{center}
\includegraphics[width=0.35\columnwidth]{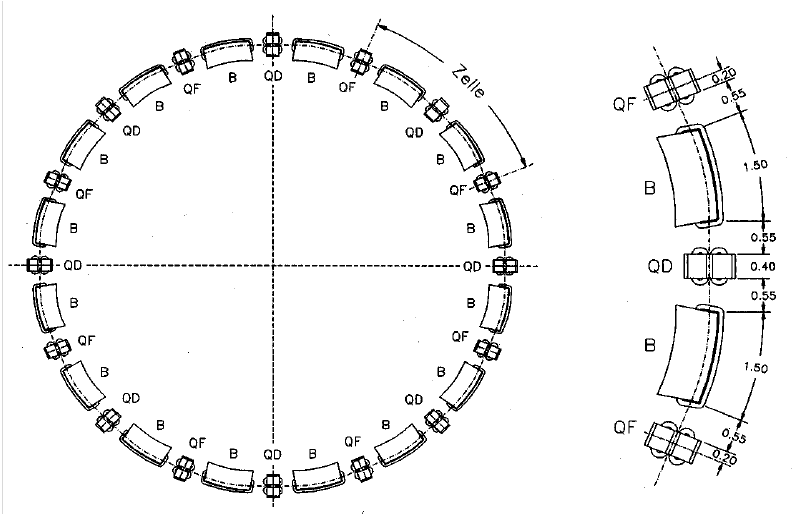}
\caption{A simple periodic chain of bending magnets and focusing/defocusing quadrupoles forming the basic structure of a storage ring.
}
\label{Willering}
\end{center}
\end{figure}

For completeness, we also include the case of a field-free drift. With $K=0$, we obtain
\[
\mathbf{M}_{\rm drift}=
\begin{pmatrix}
1          &     s  \\
0          &     1
\end{pmatrix}.
\]
This matrix formalism allows us  to combine the elements of a storage ring in an elegant way and so it is straightforward to calculate the particle trajectories. As an example, we consider the simple case of an alternating focusing and defocusing lattice, a so-called FODO lattice \cite{WiIle}; see Fig.~\ref{Willering}.

\begin{figure}[h]
\begin{center}
\includegraphics[width=0.7\columnwidth]{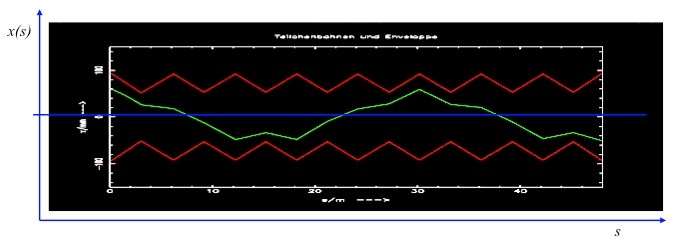}
\caption{Calculated particle trajectory in a simple storage ring%
}
\label{track_1}
\end{center}
\end{figure}

As we know the properties of each and every element in the accelerator, we can construct the corresponding matrices and calculate step by step the amplitude and angle of a single particle's trajectory around the ring.
Even more conveniently, we can multiply out the different matrices and, given initial conditions $x_{0}$ and $x'_{0}$, obtain directly the trajectory at any location in the ring:
\begin{equation}
\mathbf{M}_{\rm total} =  \mathbf{M}_{\rm foc} \cdot \mathbf{M}_{\rm drift}\cdot \mathbf{M}_{\rm dipole}\cdot \mathbf{M}_{\rm drift} \cdot \mathbf{M}_{\rm defoc}\cdots.
\end{equation}
The trajectory thus obtained is shown schematically in Fig.~\ref{track_1}.

We emphasize the following facts  in this context.
\begin{itemize}
\item At each moment, or in each lattice element, the trajectory is a part of a harmonic oscillation.
\item However, due to the different restoring or defocusing forces, the solution will look different at each location.
\item In the linear approximation that we make use of in this context, all particles experience the same external fields, and their trajectories will differ only because of their different initial conditions.
\item There seems to be an overall oscillation in both transverse planes while the particle is travelling around the ring. Its amplitude stays well within the boundaries set by the vacuum chamber, and its frequency in the example of Fig.~\ref{track_1} is roughly 1.4 transverse oscillations per revolution, which corresponds to the eigenfrequency of the particle under the influence of the external fields.
\end{itemize}

\begin{figure}[h]
\begin{center}
\includegraphics[width=0.55\columnwidth]{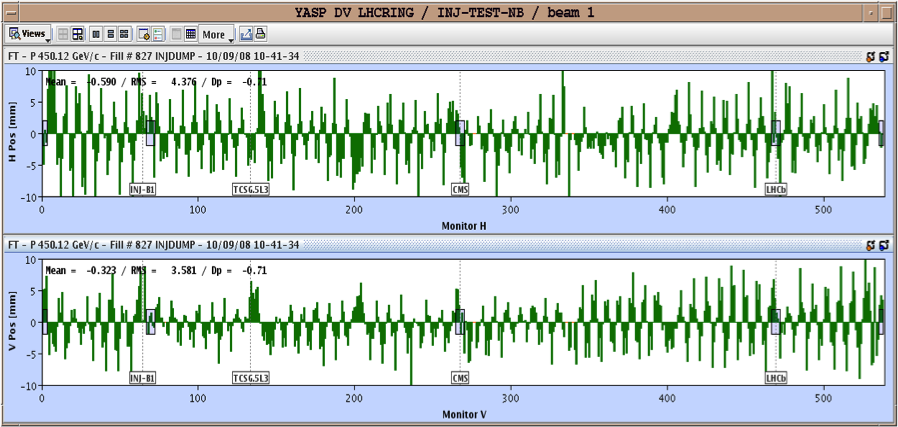}
\caption{Measured orbit in LHC during the commissioning of the machine%
}
\label{LHC_orbit}
\end{center}
\end{figure}

Coming closer to a real existing machine, we show in Fig.~\ref{LHC_orbit} the orbit measured during one of the first injections into the LHC storage ring. The horizontal oscillations are plotted in the upper half of the figure and the vertical oscillations in the lower half, on a scale of $\pm 10$~mm. Each histogram bar indicates the value recorded by a beam position monitor at a certain location in the ring,  and the orbit oscillations are clearly visible. By counting (or, better, fitting) the number of oscillations in both transverse planes, we obtain values of
\begin{equation}
Q_{x}=64.31, \qquad
Q_{y}=59.32.
\end{equation}
These values, which describe the eigenfrequencies of the particles, are called the horizontal and vertical \emph{tune}, respectively. Knowing the revolution frequency,  we can easily calculate the transverse oscillation frequencies, which for this type of machine usually lie in the range of~kHz.

\begin{figure}[h]
\begin{center}
\includegraphics[width=0.55\columnwidth]{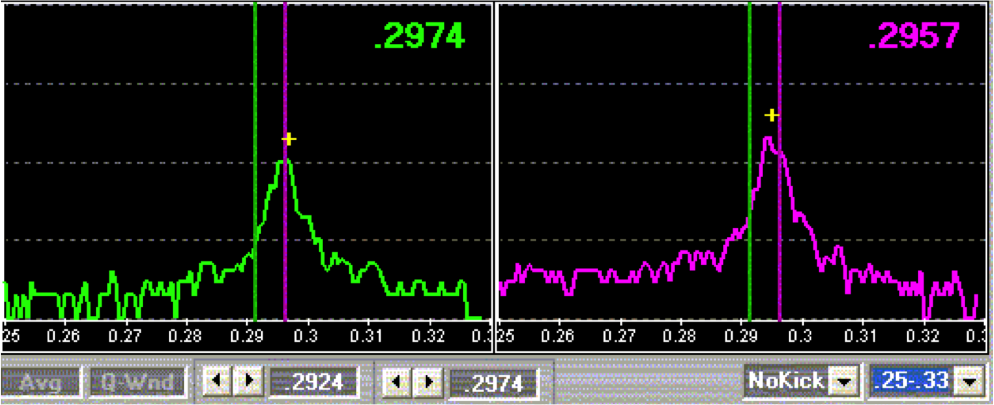}
\caption{Tune signal of a proton storage ring (HERA-p)%
}
\label{HERA_tune}
\end{center}
\end{figure}

As the tune characterizes the particle oscillations under the influence of all external fields, it is one of the most important parameters of the storage ring. Therefore it is usually  displayed and controlled at all times by the control system of such a machine. As an example, Fig.~\ref{HERA_tune} shows  the tune diagram of the HERA proton ring \cite{HERA}; it was obtained via a Fourier analysis of the spectrum measured from the signal of the complete particle ensemble. The peaks indicate the two tunes in the horizontal and vertical planes of the machine, and in a sufficiently linear machine a fairly  narrow spectrum is obtained.

\begin{figure}[h]
\begin{center}
\includegraphics[width=0.55\columnwidth]{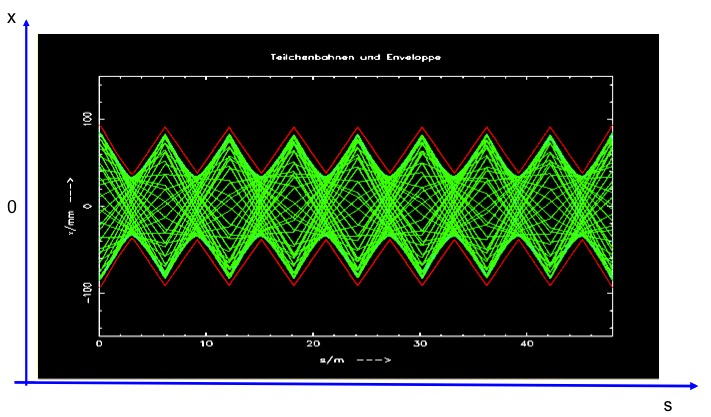}
\caption{Many single-particle trajectories together form a pattern that corresponds to the beam size in the ring
}
\label{track_n}
\end{center}
\end{figure}

Briefly referring back to Fig.~\ref{track_1}, the question is what the trajectory of the particle will look  like for the second turn, or the third, or after an arbitrary number of turns. Now, as we are dealing with a circular machine, the amplitude and angle, $x$ and $x'$, at the end of the first turn will be the initial conditions for the second turn, and so on. After many turns the overlapping trajectories begin to form a pattern, such as that in Fig.~\ref{track_n},  which indeed looks like a beam having here and there a larger and smaller beam size but still remaining well-defined in its amplitude by the external focusing forces.

To make a long story short \cite{floquet}, a mathematical function, which we call $\beta$ or amplitude function, can be defined that describes the envelope of the single-particle trajectories.
With this new variable, we can rewrite the equation for the amplitude of a particle's transverse oscillations as
\begin{equation}
x(s)=\sqrt{\epsilon}\sqrt{\beta(s)}\cos(\psi(s) + \phi),
\label{epsbeta}
\end{equation}
where $\psi$ is the phase of the oscillation, $\phi$ is its initial condition, and $\epsilon$ is a characteristic parameter of the single particle or, if we are considering a complete beam, of the ensemble of particles. Indeed, $\epsilon$ describes the space occupied by the particle in the transverse (here simplified two-dimensional) $(x, x')$ phase space. More specifically, the area  in $(x, x')$ space that is covered by the particle is given by
\begin{equation}
    A=\pi \cdot \epsilon,
\end{equation}
and, as long as we consider conservative forces acting on the particle, this area is constant according to Liouville's theorem. Here we take these facts as given, but we point out that, as a direct consequence, the so-called emittance $\epsilon$ cannot be influenced by whatever external fields are applied; it is a property of the beam, and we have to take it as given and handle it with care.

\begin{figure}[h]
\begin{center}
\includegraphics[width=0.35\columnwidth]{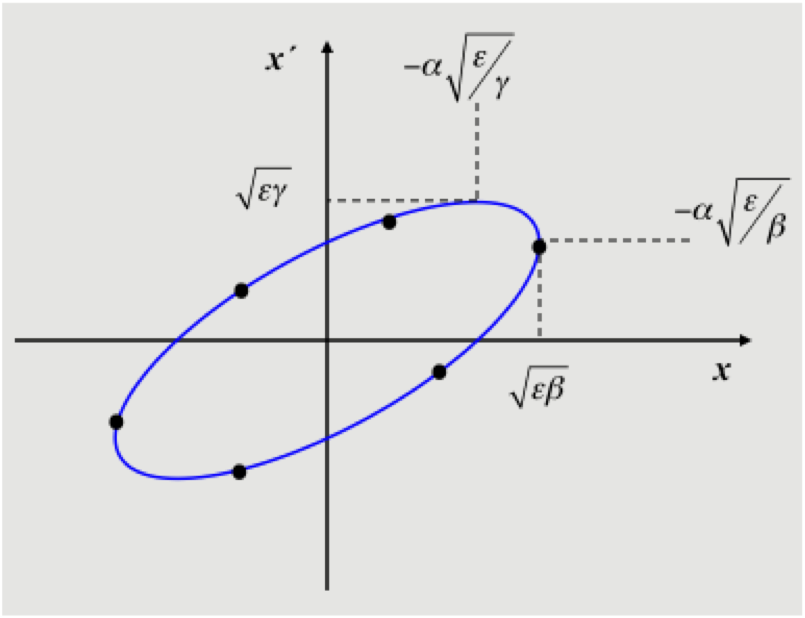}
\caption{Ellipse in $(x, x')$ phase space%
}
\label{Ellipse}
\end{center}
\end{figure}

To be more precise, and following the usual textbook treatment of accelerators, we can draw in phase space the ellipse of the particle's transverse motion; see, for example, Fig.~\ref{Ellipse}. While the shape and orientation are determined by the optics function $\beta $ and its derivative, $\alpha=-\frac{1}{2}\beta'$, the area covered is constant.

\begin{figure}[h]
\begin{center}
\includegraphics[width=0.4\columnwidth]{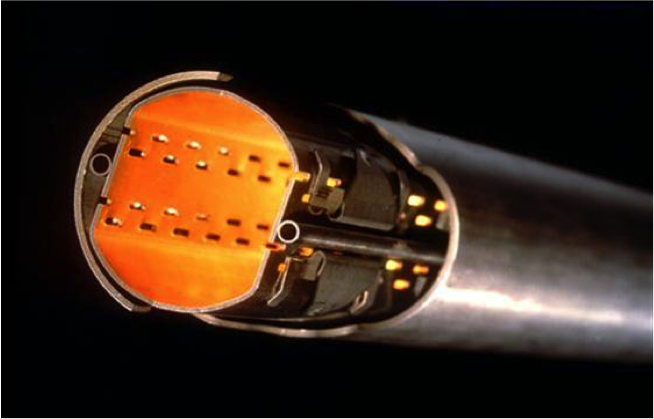}
\caption{The LHC vacuum chamber with the beam screen to shield the superconducting magnet bore from synchrotron radiation.
}
\label{Beam_screen}
\end{center}
\end{figure}

Let us talk a bit more about the beam as an ensemble of many (typically $10^{11}$) particles. Referring to Eq.~(\ref{epsbeta}), at a given position in the ring the beam size is defined by the emittance $\epsilon$ and the function $\beta$. Thus, at a certain moment in time the cosine term in (\ref{epsbeta}) will be 1 and the trajectory amplitude will reach its maximum value. Now, if we consider a particle at one standard deviation (sigma) of the transverse density distribution, then using  the emittance of this reference particle we can calculate the size of the complete beam, in the sense that the complete area (within one sigma) of all particles in $(x, x')$ phase space is surrounded (and thus defined) by our one-sigma candidate.
Thus the value $\sqrt{\epsilon \cdot \beta(s)}$ will define the one-sigma beam size in the transverse plane.
As an example, we use the values for the LHC proton beam:
in the periodic pattern of the arc, the beta function is $\beta=180$~m and the emittance at flat-top energy is roughly $\epsilon=5 \times 10^{-10}$~rad\,m.
The resulting typical beam size is therefore 0.3~mm. Now, clearly  we would not design a vacuum aperture of the machine based on a one-sigma beam size; typically, an aperture requirement corresponding to 12$\sigma$ is a good rule to guarantee a sufficient aperture, allowing for tolerances from magnet misalignments, optics errors and operational flexibility. In Fig.~\ref{Beam_screen} the LHC vacuum chamber is shown, including the beam screen used to protect the cold bore from synchrotron radiation; it corresponds to a minimum beam size of $18\sigma$.

\section {Particle colliders}
\subsection {Limit V: Fixed-target collider}

\begin{figure}[h]
\begin{center}
\includegraphics[width=0.4\columnwidth]{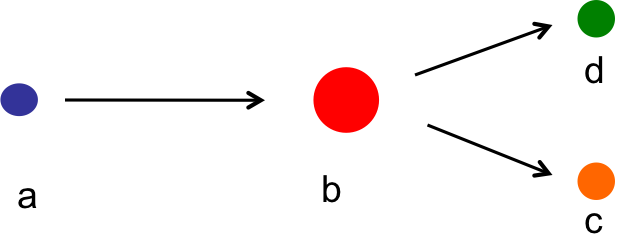}
\caption{Schematic diagram of fixed-target collider}
\label{fixed_target_scheme}
\end{center}
\end{figure}

The easiest way to perform physics experiments with particle accelerators is to bang the accelerated beam onto a target and analyse the resulting events.
While nowadays in high-energy physics we do not apply this technique any more, it still plays an essential role in the regime of atomic and nuclear physics experiments.
The advantage is that it is quite simple once the accelerator has been designed and built, and the  particles produced are easily separated due to the kinematics of the reaction. The situation is illustrated in Fig.~\ref{fixed_target_scheme}. The particle `a' that is produced and accelerated in the machine is directed onto the particle `b', which is at rest in the laboratory frame. The  particles produced from this collision are labelled `c' and `d' in this example.

While the set-up of such a scheme is quite simple, it is worth taking a closer look at the available energy in the centre-of-mass system.
The relativistic overall energy is given by
\begin{equation}
E^{2}=p^{2}c^{2}+m^{2}c^{4},
\end{equation}
which holds for a single particle but is equally valid for an ensemble of particles. Most important, the rest energy of the particle ensemble is constant (and is sometimes called the `invariant mass of the system').

Considering the system of two particles colliding, we can write
\begin{equation}
(E^{\rm cm}_{\rm a}+E^{\rm cm}_{\rm b})^{2}-(p^{\rm cm}_{\rm a}+p^{\rm cm}_{\rm b})^{2} c^{2}=(E^{\rm lab}_{\rm a}+E^{\rm lab}_{\rm b})^{2}-(p^{\rm lab}_{\rm a}+p^{\rm lab}_{\rm b})^{2} c^{2}.
\end{equation}
In the frame of the centre-of-mass system we get, by definition,
\begin{equation}
p^{\rm cm}_{\rm a}+p^{\rm cm}_{\rm b}=0,
\end{equation}
while in the laboratory frame where particle `b' is at rest we have simply
\begin{equation}
p^{\rm lab}_{\rm b}=0.
\end{equation}
The equation for the invariant mass therefore simplifies to
\begin{equation}
W^{2}=(E^{\rm cm}_{\rm a}+E^{\rm cm}_{\rm b})^{2}=(E^{\rm lab}_{\rm a}+m_{\rm b} \cdot c^{2})^{2}-(p^{\rm lab}_{\rm a}\cdot c^{2}).
\end{equation}
In other words, the energy that is available in the centre-of-mass system depends on the square root of the energy of particle `a', which is the energy provided by the particle accelerator:
\begin{equation}
W \approx \sqrt{2E^{\rm lab}_{\rm a} \cdot m_{\rm b}\cdot c^{2}}
\end{equation}
---a quite unsatisfactory situation!

\begin{figure}[h]
\begin{center}
\includegraphics[width=0.5\columnwidth]{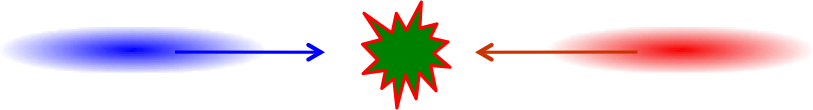}
\caption{Schematic diagram of the collision of two particles with equal energy%
}
\label{coll_target_scheme}
\end{center}
\end{figure}

To meet the demand for higher and higher energies in particle collisions, the design of modern high-energy accelerators has naturally concentrated on the development of particle colliders, where two counter-rotating beams are brought into collision at one or several interaction points (Fig.~\ref{coll_target_scheme}).

\begin{figure}[h]
\begin{center}
\includegraphics[width=0.5599999999999999\columnwidth]{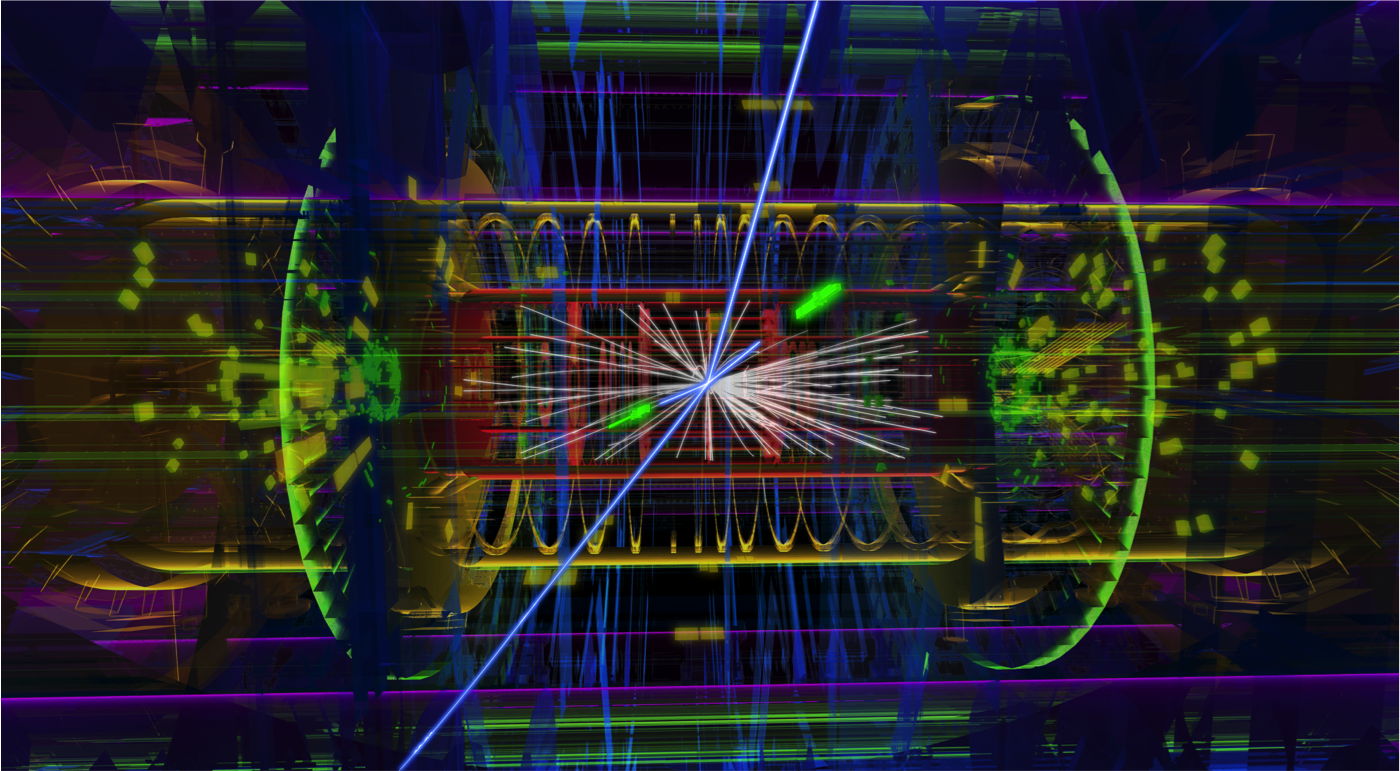}
\caption{`Typical' event observed in a collider ring---a Higgs particle measured in the ATLAS detector  %
}
\label{Higgs_event}
\end{center}
\end{figure}

If we calculate the available energy in the centre-of-mass system for the case of two colliding beams of identical particles, we get
\begin{equation}
(p^{\rm cm}_{\rm a}+p^{\rm cm}_{\rm b})^{2}=0
\end{equation}
and, by symmetry, also
\begin{equation}
(p^{\rm lab}_{\rm a}+p^{\rm lab}_{\rm b})^{2}=0.
\end{equation}
So the full energy delivered to the particles in the accelerator is available during the collision process:
\begin{equation}
W= E^{\rm lab}_{\rm a}+E^{\rm lab}_{\rm b} =2\, E^{\rm lab}_{\rm a}.
\end{equation}
A `typical' example of a high-energy physics event in such a collider is shown in Fig.~\ref{Higgs_event}.

\subsection {Limit VI: The unavoidable particle detectors}
\begin{figure}[h]
\begin{center}
\includegraphics[width=0.5599999999999999\columnwidth]{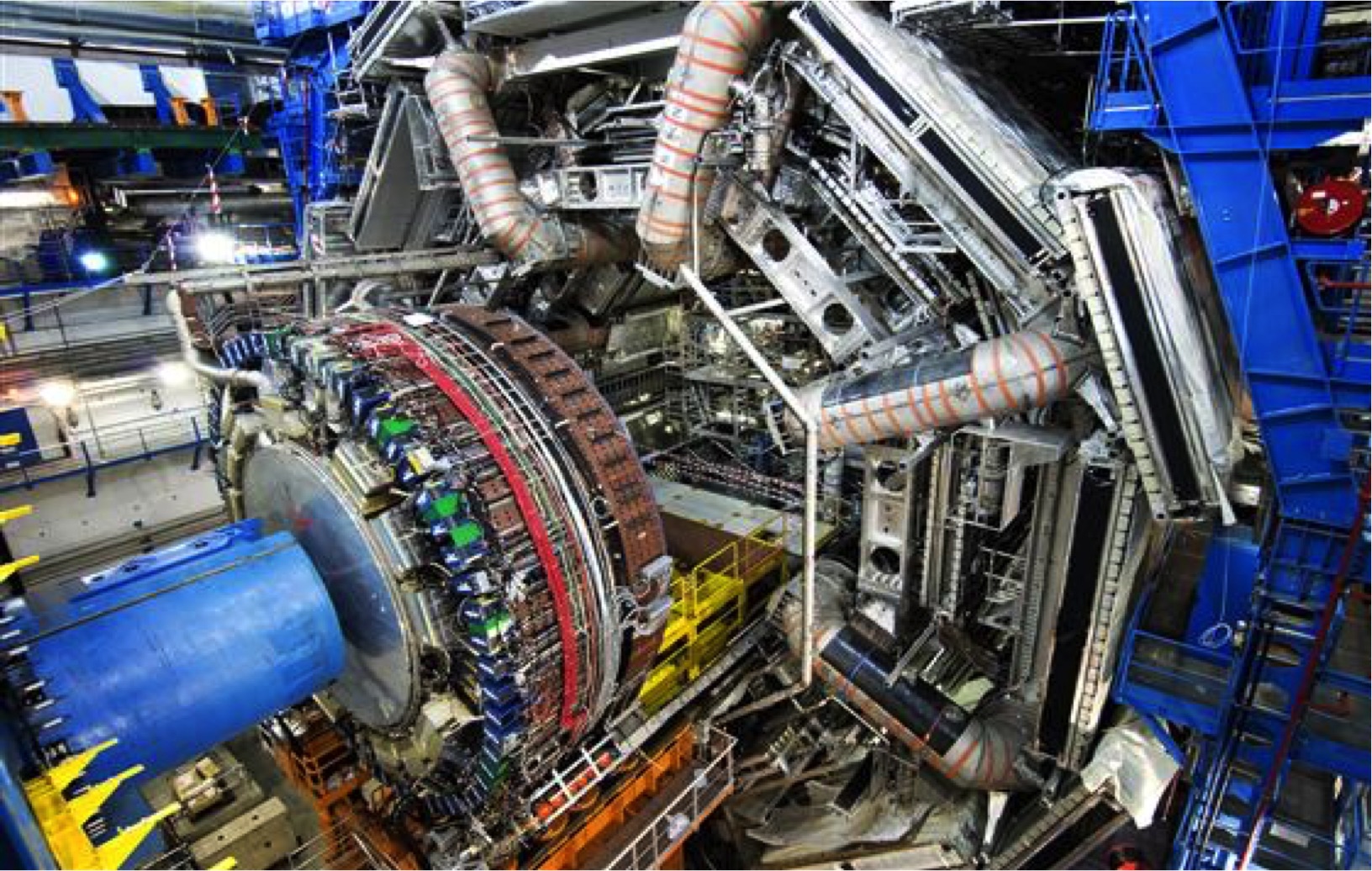}
\caption{ATLAS detector at the LHC, which is 46~m in length and has an overall weight of 7000~t%
}
\label{Atlas}
\end{center}
\end{figure}

While it is quite clear that a particle collider ring is a magnificent machine in the quest for higher energies, there is a small problem involved, namely the `particle detector'. In the arc of the storage ring we can usually  find a nice pattern of magnets providing us with a well-defined beam size, expressed as the beta function. However, special care has to be taken when our colleagues from high-energy physics  wish to install a particle detector. Especially when working at the energy frontier, just like for the accelerators, these devices tend to expand considerably in size with the energies required.
In  Fig.~\ref{Atlas}, the largest particle detector installed in a storage ring is shown as an impressive example: the ATLAS detector at the LHC.

The storage ring has to be designed to provide the space needed for the detector hardware and at the same time create the smallest achievable beam spots at the collision point, which is usually right in the centre of the detector. Unfortunately these requirements are a bit contradictory. The equation for the luminosity of a particle collider depends on the stored beam currents and the transverse spot size of the colliding beams at the interaction point (IP):
\begin{equation}
L=\frac{1}{4 \pi e^{2}f_{0}b}\cdot \frac{I_{1}I_{2}}{\sigma_{x}^{*}\sigma_{y}^{*}}.
\end{equation}
At the same time, however, the beta function in a symmetric drift grows quadratically as a function of the distance between the beam waist and the first focusing element, i.e.
\begin{equation}
\beta(s)=\beta^{*}+\frac {s^{2}}{\beta^{*}}.
\end{equation}
The smaller the beam at the IP, the faster it will grow until we can apply---outside of the detector region---the first quadrupole lenses. As a consequence, this trend sets critical limits on the achievable quadrupole aperture or, for a given aperture, the achievable quadrupole gradient. The focusing lenses right before and after the IP, being placed as closed as possible to the detector, are generally the most critical and most expensive magnets in the machine, and their aperture requirement ultimately  determines the luminosity that can be delivered by the storage ring.

For the experts we would like to add that even if the bare aperture requirement can be fulfilled, the resulting chromaticity that is created in the mini-beta insertion and the sextupole strengths that are needed to correct for it usually pose the next limit that we will face.

\subsection {Limit VII: The relative rareness of Nobel prize-winning reactions}

\begin{figure}[h]
\begin{center}
\includegraphics[width=0.45\columnwidth]{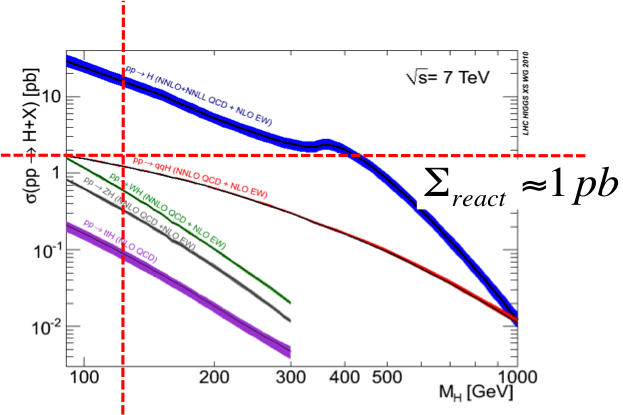}
\caption{Cross-section of the Higgs particle for different production processes,  courtesy of the CMS collaboration%
}
\label{Higgs_cross}
\end{center}
\end{figure}

The rate of events produced in a particle collision process depends not only on the performance of the colliding beams but first and foremost on the probability of creating such an event, the so-called cross-section of the process.
In the case of the Higgs particle, which is without doubt the highlight of LHC Run~1, the overall cross-section is displayed in Fig.~\ref{Higgs_cross}.

Without going into details, we can state that the cross-section for Higgs production is on the order of
\begin{equation}
\Sigma_{\rm react} \simeq 1~\mathrm{pb}.
\end{equation}
During the three years of LHC Run~1, i.e.\ the period 2011--2013, an overall luminosity of
\begin{equation}
\int L\,\rmd t = 25~\mathrm{fb}^{-1}
\end{equation}
was accumulated.

Combining these two numbers using the fact that the event rate of a reaction is $R=L \cdot \Sigma_{\rm react}$, we get a total number of `some thousand' Higgs particles produced---for a Nobel prize-winning event just at the edge of reliable statistics.
Therefore, the particle colliders have to be optimized not only for the highest achievable energies but  also for maximum stored beam currents and small spot sizes at the interaction points so as to optimize the luminosity of the machine.

\subsection {Limit VIII: The luminosity of a collider ring}
Following the arguments above, the design goal here is to prepare, accelerate and store two counter-rotating particle beams in order to profit best from the energy of the two beams during the collision process. Still, there is a price to pay: unlike in fixed-target experiments, where the `particle' density of the target material is extremely high, in the case of two colliding beams the event rate is basically determined by the transverse particle density that can be achieved at the IP.
Assuming Gaussian density distributions in both transverse planes, the performance of such a collider is described by the luminosity
\begin{equation}
L=\frac{1}{4 \pi e^{2}f_{0}b}\cdot \frac{I_{1}I_{2}}{\sigma_{x}^{*}\sigma_{y}^{*}}.
\end{equation}
While the revolution frequency $f_{0}$ and the bunch number $b$ are ultimately determined by the size of the machine, the stored beam currents $I_{1} $ and $I_{2}$ and the beam sizes $\sigma_{x}^{*}$ and $\sigma_{y}^{*}$ at the IP  have their own limitations.

The most serious limitation comes from the beam--beam interaction itself. During the collision process, individual particles of the counter-rotating  bunches feel the space charge of the opposing bunch. In the case of a proton--proton collider, this strong field acts like a defocusing lens, and has a strong impact on the tune of the bunches \cite{Herr}.

\begin{figure}[h]
\begin{center}
\includegraphics[width=0.4\columnwidth]{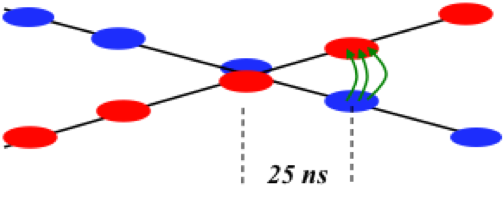}
\caption{Schematic view of the beam--beam interaction during the crossing of bunch trains%
}
\label{bb_trains}
\end{center}
\end{figure}

\begin{figure}[h]
\begin{center}
\includegraphics[width=0.6\columnwidth]{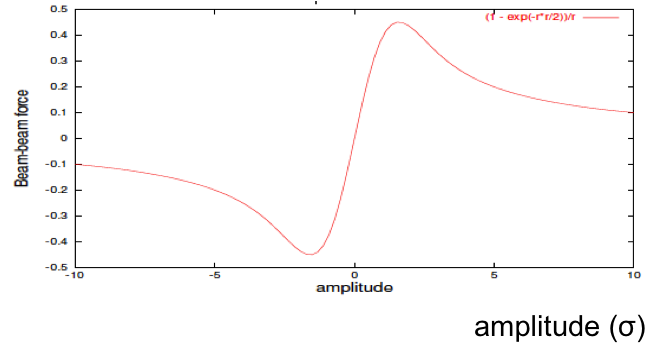}
\caption{Beam--beam force as a function of the transverse distance of the particle to the centre of the opposing bunch.
}
\label{bb_force}
\end{center}
\end{figure}

In  Fig.~\ref{bb_trains} the situation is shown schematically. Two bunch trains collide at the IP, and during the collision process a direct beam--beam effect is observed. In addition to that, before and after the actual collision, long-range forces exist between the bunches that have a nonlinear component; see Fig.~\ref{bb_force}. As a consequence \cite{Herr}, the tune of the beams is not only shifted with respect to the natural tune of the machine but also spread out, as different particles inside the bunches see different contributions from the beam--beam interaction.

\begin{figure}[h]
\begin{center}
\includegraphics[width=0.5\columnwidth]{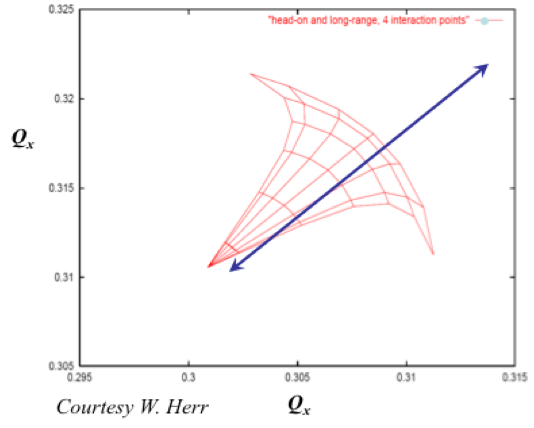}
\caption{Calculated tune shift due to beam--beam interaction in the LHC%
}
\label{bb_working_diagram}
\end{center}
\end{figure}

Therefore, in the tune diagram, we no longer obtain a single spot representing the ensemble of  particles, but rather a large array that depends in shape, size and orientation on the particle densities, the distance of the bunches at the long-range encounters, and the single-bunch intensities. The effect has been calculated for the LHC and is displayed in Fig.~\ref{bb_working_diagram}.

In a number of cases a useful approximation can be applied, as for distances of about 1--2$\sigma$, the beam--beam force in Fig.~\ref{bb_force} can be linearized and acts like a quadrupole lens. Accordingly, a tune shift can be calculated to characterize the strength of the beam--beam effect in a collider.
Given the parameters described above, and introducing the classical particle radius $r_{\rm p}$, the amplitude function $\beta^{*}$ at the IP and the Lorentz factor $\gamma$, we can express the tune shift due to the linearized beam--beam effect as
\begin{equation}
\Delta Q_{y}=\frac{\beta_{y}^{*}\cdot r_{\rm p} \cdot N_{\rm p}}{2 \pi \, \gamma (\sigma_{x}+\sigma_{y})\sigma_{y}}.
\end{equation}
In the case of the LHC, the design value of the beam--beam tune shift is $\Delta Q=  0.0033 $, and in daily operation the machine is optimized to run close to this value, which places the ultimate limit on achievable bunch intensities in the collider.

\section {Lepton colliders}
\subsection {Limit IX: Synchrotron light---the drawback of electron storage rings}
In proton or heavy-ion storage rings, the design can  more or less follow the rules discussed above.
But the situation changes drastically as the particles become more and more relativistic. Bent on a circular path, electrons in particular will radiate an intense light, the so-called synchrotron radiation, which will have a strong influence on the beam parameters as well as on the design of the machine.

\begin{figure}[h]
\begin{center}
\includegraphics[width=0.5\columnwidth]{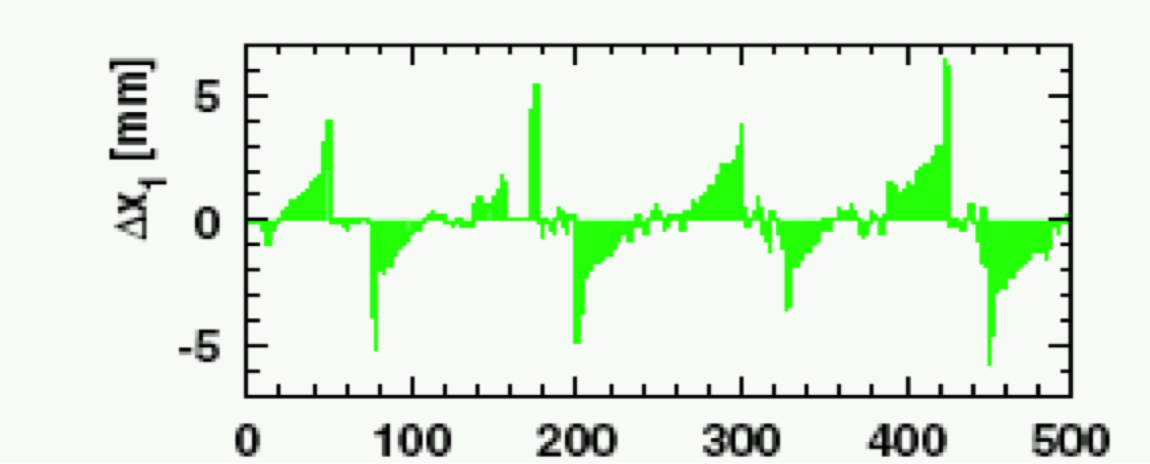}
\caption{Measured horizontal orbit of the LEP electron beam; due to synchrotron radiation losses, the particle orbit is shifted towards the inner side of the ring in each arc.%
}
\label{saw_tooth}
\end{center}
\end{figure}

Summarizing the situation briefly here, the power loss due to synchrotron radiation depends on the bending radius and the energy of the particle beam:
\begin{equation}
P_{\rm s}=\frac{2}{3}\alpha \hbar c^{2} \frac{\gamma^{4}}{\rho^2},
\label{syli}
\end{equation}
where $\alpha$ represents the fine structure constant and $\rho$ the bending radius in the dipole magnets of the ring.
As a consequence, the particles will lose energy turn by turn.
To compensate for these losses, RF power has to be supplied to the beam at any moment. An example that illustrates the problem nicely is shown in Fig.~\ref{saw_tooth}. It plots the horizontal orbit of the former  Large Electron--Positron Collider (LEP) storage ring. The electrons, travelling from  right to left in the plot, lose a considerable amount of energy in each arc and hence deviate from the ideal orbit towards the inner side of the ring. The effect on the orbit is large: up to 5~mm orbit deviation was  observed in the example of Fig.~\ref{saw_tooth}. In order to compensate for these losses, four RF stations were installed in the straight sections of the ring to supply the necessary power.

The strong dependence of the synchrotron radiation losses on the relativistic $\gamma$ factor sets severe limits on the beam energy that can be carried in a storage ring of a given size. The push for ever higher energies means either that storage rings even  larger than LEP need to be designed or,  to avoid synchrotron radiation, linear accelerating structures should be developed.

\begin{figure}[h]
\begin{center}
\includegraphics[width=0.45\columnwidth]{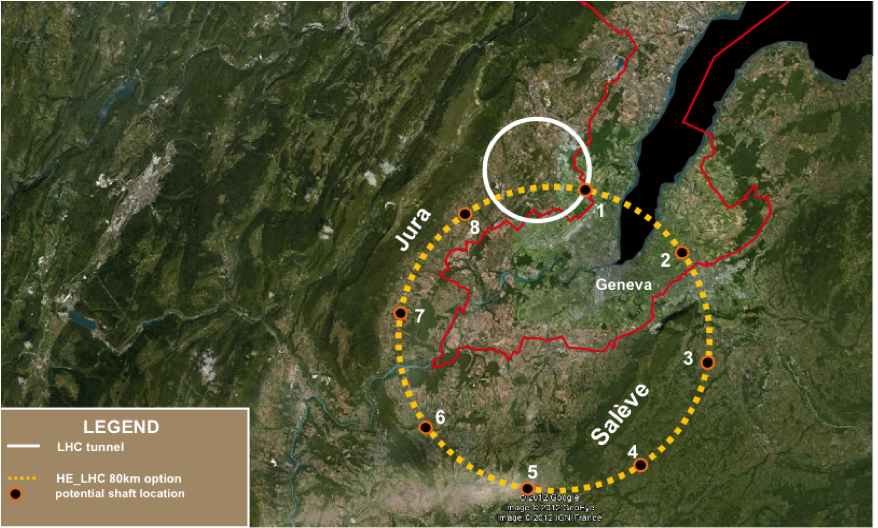}
\caption{Schematic view of a possible 100~km FCC design in the Geneva region%
}
\label{FCC}
\end{center}
\end{figure}

Currently, the next generation of particle colliders is being studied \cite{TLEP}. The ring design of the future circular collider (FCC) foresees a 100~km ring to carry electrons (and positrons) of up to 175~GeV energy.
The size of this storage ring is far beyond the dimensions of anything that has been designed up to now. A  sketch of the machine layout is given in  Fig.~\ref{FCC}, where the yellow dashed circle delineates the  100~km ring and the white circle represents the little LHC machine.

For the maximum projected electron energy of $E=175$~GeV, synchrotron radiation would cause an energy loss of 8.6~GeV, or an overall power of 47~MW of the radiated light at full beam intensity.

\subsection {Limit X: Acceleration gradients in linear structures}
As far as lepton beams are concerned, ring colliders suffer from the severe limitation caused by synchrotron radiation losses, and at a certain point the construction of such large facilities would not seem reasonable any more.
To avoid the problem of synchrotron radiation, linear structures that were discussed earlier and used in the infancy of particle accelerators have become in vogue again.
Still, the advantage of circular colliders cannot be completely ignored:
even with a modest acceleration gradient in the RF structures, the particles will get turn by turn a certain boost in energy and will at some point reach the desired flat-top energy in the ring.

\begin{figure}[h]
\begin{center}
\includegraphics[width=0.8\columnwidth]{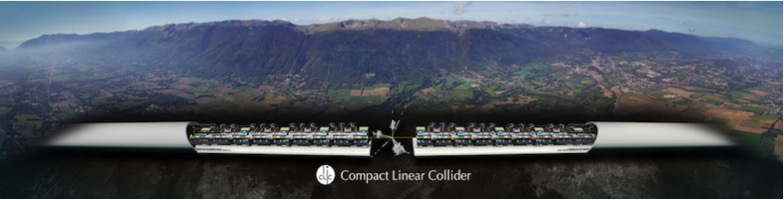}
\caption{Proposed location of the CLIC linear collider along the Jura mountains in the Geneva region%
}
\label{CLIC_photo}
\end{center}
\end{figure}

In a linear accelerator, this kind of repetitive acceleration is by design not possible; within a single pass through the machine, the particles will have to be accelerated to full energy. In order to keep the structure compact, the highest acceleration gradients will therefore be needed. One of the most prominent designs proposed for a possible future collider is the CLIC design \cite{CLIC}. Within one passage through the 25~km long accelerator, the electron  beam will get up to 3~TeV, and the same is true for the opposing positron beam.
An artist's rendering of this machine is shown in Fig.~\ref{CLIC_photo}.

The main parameters of the CLIC design are listed in Table~\ref{clic_param}. The accelerating gradient, i.e.\ the energy gain per meter, is especially to be emphasized; it has been pushed to the maximum value that is technically feasible, and the limit is ultimately due to the breakdown of the electric field in the accelerating structure.
\begin{table}[h]
 \centering
 \caption{Main parameters of the CLIC study}
\label{clic_param}
  \begin{tabular}{lcc}
\hline
\hline
 &    {\bf 500 GeV}                     & {\bf 3 TeV}                          \\
\hline
Site length                                 & 13 km                             &  48 km                         \\
Loaded acceleration gradient (MV\,m$^{-1}$)  &                     \multicolumn{2}{c} {12}                        \\
Beam power per beam (MW)            & 4.9                                 &  14                                \\
Bunch charge ($10^{9} $ e+/e)   & 6.8                                 &  3.7                                \\
Horizontal/vertical normalized emittance ($10^{-6}/10^{-9}$~m)     &       2.4/25   &  0.66/20                        \\
Beta function (mm)                   &    \multicolumn{2}{c} {10/0.07}                              \\
Beam size at IP: horizontal/vertical (nm)      &   \multicolumn{2}{c} {45/1}                                     \\\\
 Luminosity (cm$^{-2}$\,s$^{-1}$)      &  $2.3 \times 10^{34} $      &  $5.9 \times 10^{34} $  \\
 \hline\hline
  \end{tabular}
\end{table}

\begin{figure}[b]
\begin{center}
\includegraphics[width=0.7\columnwidth]{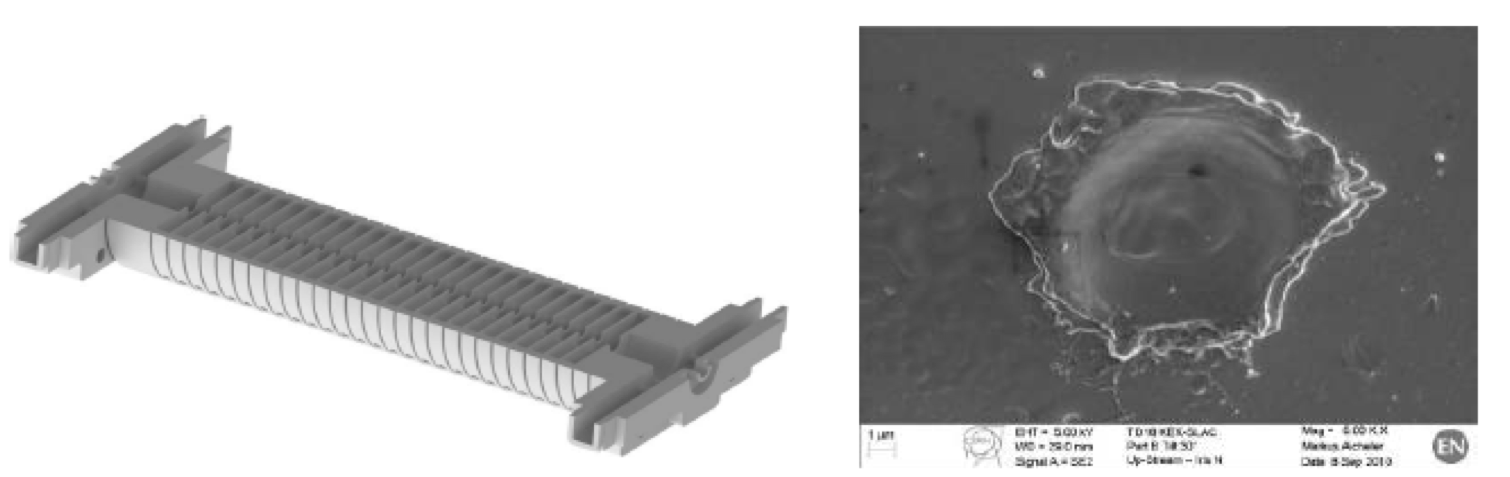}
\caption{Accelerating structure of the CLIC test facility CTF3; the electron microscope photo on the right shows the damage to the surface due to discharges in the module.%
}
\label{CLIC_structure}
\end{center}
\end{figure}

A picture of such a CLIC-type structure is shown in Fig.~\ref{CLIC_structure}. On the right-hand side is an electron microscope photo of the surface after a voltage breakdown. At the spot of the sparking, a little crater can be seen, indicating possible damage to the surface and, as a consequence,  deterioration of the achievable gradient which has to be avoided under all circumstances \cite{Palaia}.
Although considerably higher than the typical values in circular machines, the gradient $E_{\rm acc}~=~100$~MV\,m$^{-1}$ in a linear machine still leads to a design of  overall length approximately  50~km for a maximum achievable energy of $E_{\max}=3$~TeV.

\section {Conclusion}
\begin{figure}[h]
\begin{center}
\includegraphics[width=0.6\columnwidth]{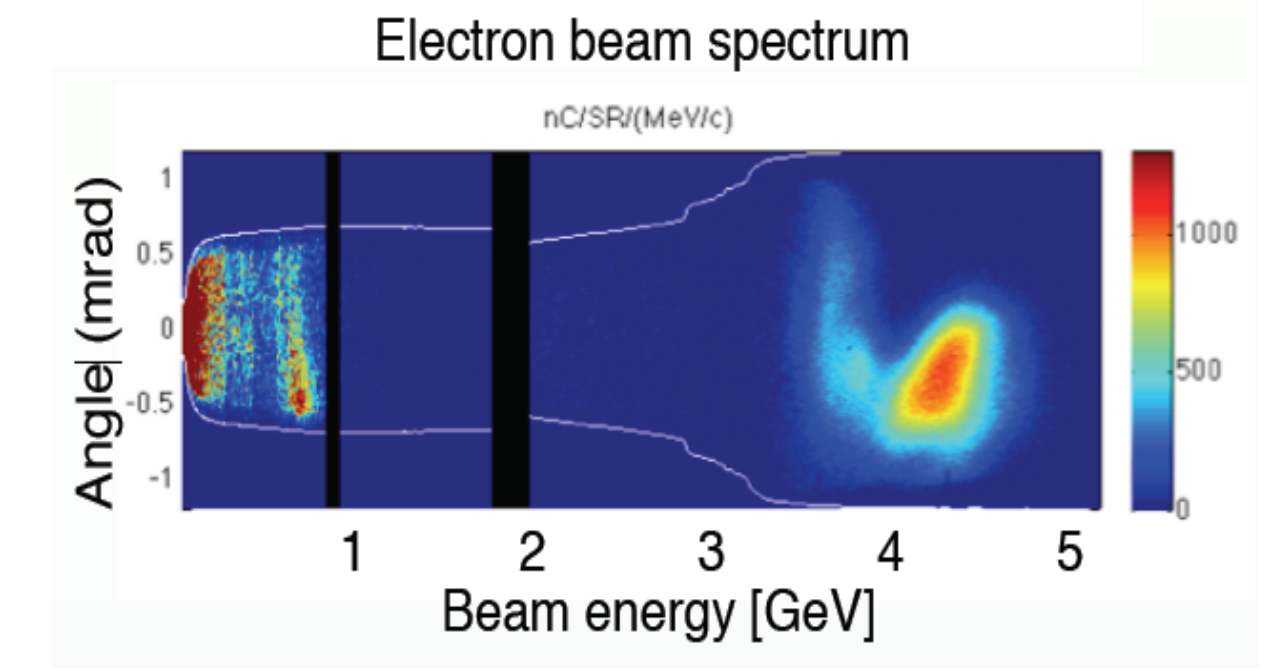}
\caption{Electron beam accelerated in the wake potential of a plasma cell; up to 4~GeV is obtained within only a few centimetres of length\cite{pwa}.%
}
\label{pwa}
\end{center}
\end{figure}

To summarize, for future lepton ring colliders (or, to be more precise, electron--positron colliders),  synchrotron radiation losses set a severe limit on the achievable beam energy; and very soon the size of the machines will become uneconomical. For a given limit in synchrotron radiation power, the dimensions of the machine would have to grow quadratically with the beam energy. Linear colliders are therefore proposed as the preferred way to go. In this case, the maximum achievable acceleration gradient is the key issue. New acceleration methods, namely plasma-based set-ups in which gradients have been observed that are much higher than those seen with conventional techniques, are a most promising concept for the design of future colliders. An impressive example is shown in Fig.~\ref{pwa}: within a plasma cell of only a few centimetres in  length, electrons are accelerated to several GeV. The gradients achievable are  orders of magnitude higher than in any conventional machine (see, e.g., Ref.~\cite{pwa}). Still, there are problems to overcome, such as issues with overall efficiency, beam quality (mainly the energy spread of the beam), and the achievable repetition rate. Nevertheless, we are convinced that this is a promising field worthy of much further study---and this is what the present school is about.


\bibliography{bibliography/converted_to_latex.bib%
}

\end{document}